\documentclass[conference]{IEEEtran}

\usepackage{graphicx}
\usepackage{float}
\usepackage[cmex10]{amsmath} 
\usepackage{multirow}
\usepackage{hhline}
\usepackage{mwe} 
\usepackage{subcaption}
\usepackage{cite}
\usepackage[numbers,sort&compress]{natbib}
\usepackage{tabularx}
\usepackage{lipsum,booktabs,siunitx}
\newcolumntype{T}[1]{S[table-format=#1,group-digits=false]}
\usepackage{xcolor}
\usepackage{colortbl}
\usepackage{lipsum}
\usepackage{booktabs}

\usepackage{array}
\usepackage{caption}
\begin{document}
\title{COVID-CLNet: COVID-19 Detection with Compressive Deep Learning Approaches
}

\author{\IEEEauthorblockN{Khalfalla Awedat\IEEEauthorrefmark{1} and Almabrok Essa\IEEEauthorrefmark{2}}
	\IEEEauthorblockA{\IEEEauthorrefmark{1}Department of Computer Science, Pacific Lutheran University, Tacoma, WA, USA\\
	\IEEEauthorrefmark{2}Department of Electrical Engineering and Computer Science, Cleveland State University, Cleveland, {OH}, USA\\
	Email:\IEEEauthorrefmark{1}awedatka@plu.edu; \IEEEauthorrefmark{2}a.essa@csuohio.edu}}
\maketitle

\begin{abstract}
One of the most serious global health threat is COVID-19 pandemic. The emphasis on improving diagnosis and increasing the diagnostic capability helps stopping its spread significantly. Therefore, to assist the radiologist or other medical professional to detect and identify the COVID-19 cases in the shortest possible time, we propose a computer-aided detection (CADe) system that uses the computed tomography (CT) scan images. This proposed boosted deep learning network (CLNet) is based on the implementation of Deep Learning (DL) networks as a complementary to the Compressive Learning (CL). We utilize our inception feature extraction technique in the measurement domain using CL to represent the data features into a new space with less dimensionality before accessing the Convolutional Neural Network. All original features have been contributed equally in the new space using a sensing matrix. Experiments performed on different compressed methods show promising results for COVID-19 detection. In addition, our novel weighted method based on different sensing matrices that used to capture boosted features demonstrates an improvement in the performance of the proposed method.   

\end{abstract} 


\section{Introduction}
\begin{figure}[t]
	\centering
	\hspace*{-.1cm}
	\includegraphics[width=.45\textwidth,height=4cm]{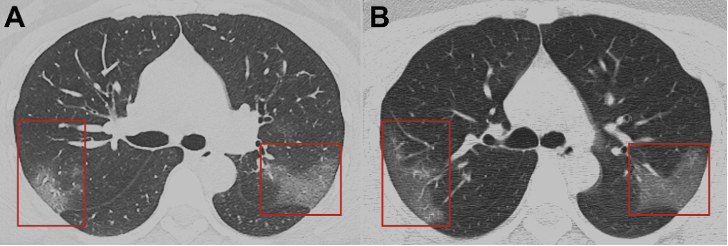}
	\caption{Unenhanced CT images. According to \cite{radiol}, image A shows multiple ground-glass opacities in bilateral lungs. While image B which obtained 3 days after follow-up shows progressive ground-glass opacities in the posterior segment of right upper lobe and apical posterior segment of left superior lobe.}
	\label{fig1}
\end{figure}

Coronavirus Disease 2019 (COVID-19) is a novel (new) virus that first identified in Wuhan, Hubei Province, China in December 2019. COVID-19 is contagious respiratory illnesses that is caused by infection with a new coronavirus (called SARS-CoV-2), which affects different people in different ways. The centers for disease control and prevention (CDC) is closely monitoring the spread of cases caused by this disease. As of the best of our knowledge while we write this article and according to the World Health Organization (WHO), more than 60 million confirmed cases globally, and more than 1 million deaths. One of the assessment or examination processes to identify COVID-19 is the chest CT, which is recommended to be used as the primary screening or diagnostic method. Therefore, the CADe systems are recommended to detect the earliest signs of ground-glass nodules in thoracic CT that are caused by this disease, which may not be detected by the radiologist or other medical professional at the early times. Fig. \ref{fig1} shows that COVID-19 causes multiple peripheral ground-glass opacities in lung that did not spare the subpleural regions, image A in the figure, which progressive produced pulmonary opacities after 3 days, image B in the figure \cite{radiol}. 

The main motivation of this research is to assist accelerating the diagnostic process and help stopping this widespread pandemic. Therefore, we introduce the CADe system that applies the advanced deep learning-based radiology image analysis methods as a complementary to the compressive learning (CL), which is based on different sensing matrices weighted strategy. This CADe system could outperform many state-of-the-art methods.

\subsection{Deep Learning}
Currently, COVID-19 radiology literature based on Deep Learning (DL) methods has been explored the use of CT images extensively and shown promising detection accuracy of COVID-19 \cite{ xu2020deep,gozes2020rapid,wang2020deep,ko2020covid}. DL approaches like machine learning which can be categorized as follows: supervised, semi-supervised, and unsupervised learning. In addition, there is another category of learning called reinforcement learning or deep reinforcement learning which are often considered to be a special case of semi supervised or sometimes unsupervised learning approaches \cite{alom2018history}.  
\subsubsection{Deep Unsupervised Learning}
Unsupervised learning grants the learning algorithm to find the structure in its input since no labels are given by discovering the hidden patterns within the input data. Often clustering, dimensionality reduction, and generative techniques are considered as unsupervised learning approaches, such as Encoders (AE), Restricted Boltzmann Machines (RBM), and the recently developed Generative Adversarial Networks (GAN).
\subsubsection{Deep Semi-supervised Learning}
Semi-supervised learning provides powerful framework for leveraging unlabeled data when labels are limited by combining that limited number of labels and a large number of unlabeled datasets (partially labeled datasets) to construct a model or classifier feature. Semi-supervised learning is between supervised and unsupervised learning. In some cases, Deep Reinforcement Learning (DRL) and GAN are used as semi-supervised learning techniques.  
\subsubsection{Deep Supervised Learning}
supervised learning is a learning technique that uses labeled data to infer the relationship between the observed data and a predetermined dependent variable. In the case of supervised DL approaches, the predetermined dependent variable has a set of inputs and corresponding outputs. After successful training, and the goal is to learn a general rule that maps inputs to outputs. There are different supervised learning approaches for deep leaning including Deep Neural Networks (DNN), Convolutional Neural Networks (CNN), Recurrent Neural Networks (RNN) including Long Short-Term Memory (LSTM), and Gated Recurrent Units (GRU). 
\subsubsection{Deep Reinforcement Learning (DRL)}
DRL is a learning technique for use in unknown environments where a system interacts with a dynamic environment in which it must perform a certain goal. In this case, there is a straightforward loss function that means do not have full access to the function you are trying to optimize. Therefore, the system is provided feedback in terms of rewards and punishments as it navigates its problem space. DLR is the appropriate way to go if the problem has a lot of parameters to be optimized.
\subsection{Compressive Sensing (CS)}
One of the very powerful signal processing techniques is the Compressive Sensing (CS), which has provided fast and efficient data acquisition in many applications. Based on the assumption that each data has a sparse representation in some basis \cite{candes2006stable, donoho2006,awedat2017sparse,awedat2017}. It has been built based on the fact that the sparse signals can be recovered with high accuracy by projecting or sensing the data into the measurement domain. The sensing data can be achieved using sensing matrix which should satisfy the incoherent, restricted isometry property (RIP) \cite{candes2005decoding}. Most of CS works have been focused on providing theories for reconstruction the original sparse data \cite{draganic2017some,pope2009compressive}. Mathematically, for a signal $x\in\Re^{N}$ is sparse $s=\|x\|_0\in\Re^{N}$(where$ \|x\|_0$ is number of nonzero entiries). Due the sparsity $x$ can be manipulate in new domain $y\in \Re^{M}$  where $M\ll{N}$ by linear system as
\begin{equation} \label{2}
y=\phi. {x}
\end{equation}
where $\phi \in\Re ^{M\times{N}}$ is sensing (or measurements) matrix. 
Decoding process focuses on finding back the sparse signal $x$ from a given measurement $y$.  For this purpose, the optimization method needs to apply as

\begin{equation} \label{2}
\min{\| y\|}_{0}  \quad subject to   \quad \phi. {x}=y     
\end{equation}  
However, this problem is NP-hard.  Instead, the reconstruction can be done use L1-minimization

\begin{equation} \label{2}
\min{\| y\|}_{1}  \quad subject to   \quad \phi. {x}=y     
\end{equation}  

Generally, the CS can be optimized in coding procedure by implement different coding matrices \cite{arjoune2018performance}, or using different optimization methods to reconstruct the original signals \cite {pope2009compressive}. While the reconstruction sparse signal is the main objective of CS, the compressed signal can be very useful in the applications that detecting certain patterns or features for classification \cite{calderbank2012finding,wright2008robust}. Moreover, in some scenarios related to information privacy, reconstruction is undesirable \cite{mohassel2017secureml}. Therefore, the Compressive Learning (CL) has been proposed \cite{calderbank2012finding,davenport2007smashed,davenport2010signal} where the system is built based on the compressed measurements without the reconstruction step. Since CL has been built based on all features of data that are combined to reduce the dimensionality, it still can be used for learning task. In other words, sine all the original features have been involved in projected domain, the new low dimension projected features can be applied to distinguish the original pattern or class \cite{awedat2020}.

This study is inspired by \cite{awedat2020}, and based on the observation that the new low dimension projected features which can be obtained by CL are great source of information to pass through advanced deep learning methods. Therefore, we propose the CL to employ our inception feature extraction technique in the measurement domain for representing the data features into a new space with less dimensionality before accessing the deep learning network. The novel scientific contributions in this paper are summarized as follows:

\begin{itemize}
	\item It introduces a computer-aided detection (CADe) system based on the boosted deep learning network (CLNet), which uses compressive learning based deep learning approach.
	\item The evaluation process of the proposed CLNet technique has been conducted on raw CT images without any preprocessing and has shown signs of high performance.	
\end{itemize}

The rest of this paper is organized as the following: The related work is covered in Section 2. The proposed approach is presented in Section 3. Section 4 discusses in details the experimental setup including data description, obtained results, detailed discussion, and the work limitations. Section 5 concludes the work and introduces the future work plans.

\section{Related work}
The effort of developing deep learning technique for diagnose COVID-19 has been gradually increased since the outbreak. To illustrate the importance of early detection and management of COVID19 patients, a detailed study has been conducted in \cite{yang2020patients, alom2020covid_mtnet}. Some literature reviews demonstrate that the multiple peripheral ground-glass opacities in lung which are caused by COVID-19 disease are appeared on CT images \ref{fig1} sometimes are not appeared on the chest X-ray (CXR) \cite{radiol,ng2020imaging}. Due to superior ability of deep learning of image classification, there are several Artificial Intelligence (AI) systems that have been proposed for COVID-19 detection based on medical imaging. While some literature reviews show that DL methods using CT images have been achieved promising detection accuracy of COVID-19, the DL based approaches also have been utilized extensively on CXR images and successfully have provided high performance. The authors in \cite{civit2020deep, narin2020automatic} proposed convolutional neural network (CNN) for the detection of coronavirus pneumonia infected patient using chest X-ray radiographs. On the other hand, the authors in \cite{butt2020deep} applied 2D and 3D deep learning to classify CT samples with COVID-19, Influenza viral pneumonia cases and no-infection cases. \cite{yang2020covid} has been built a database of COVID-19 based on CT images. The diagnosis method was based on concatenating both lung masks and lesion masks. \cite{jaiswal2020classification} proposed DenseNet201 based deep transfer learning (DTL) to classify the patients as COVID infected. There are many studies have been proposed to diagnosis coronavirus using deep learning \cite{xu2020deep,ko2020covid,ozturk2020automated,wang2020deep}. According to various studies presented in the literature, the learning framework has been applied on medical images with different approaches for preprocessing methods. In our study, by taking the advantage of CL that the images can be represented in new domains with small number of features, and then combining these features would be useful to improve the diagnosis accuracy.

\section{Proposed Method}
\begin{figure}[hbt!]
    \centering
    \captionsetup{margin=.3cm}
    \includegraphics[width=8.7cm,height=21.75cm]{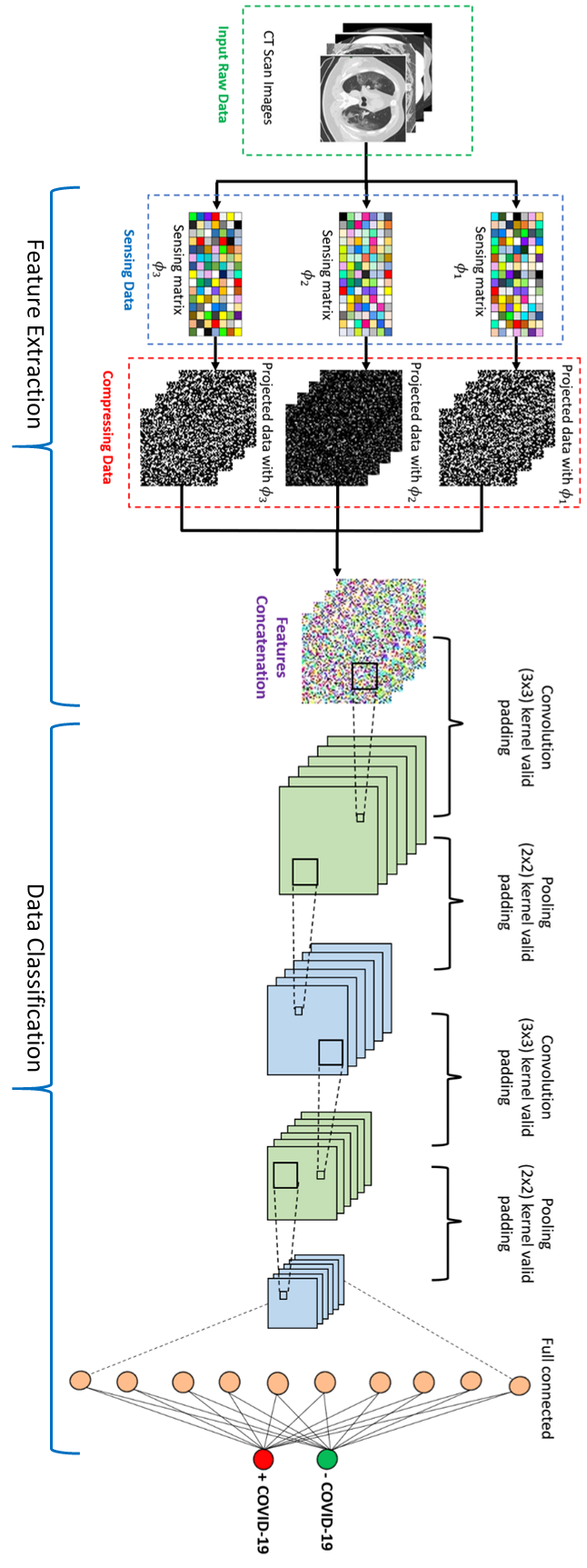}
    \caption{Block diagram of the proposed $COVID-CLNet$ method for $COVID-19$ detection.}
    \label{fig2}
\end{figure}

In this implementation, to investigate the appearance of coronavirus (COVID-19) on CT images, we have utilized our inception feature extraction network based on the compressing learning (CL) to represent the data features into a new space with less dimensionality before accessing the advanced deep learning network. The end-to-end training pipeline of the proposed CLNet is shown in Fig. \ref{fig2}. 	

\subsection{Feature Extraction}
\begin{figure}[h]
	\centering
	\hspace*{-.5cm}
	\includegraphics[width=.47\textwidth,height=5.07cm]{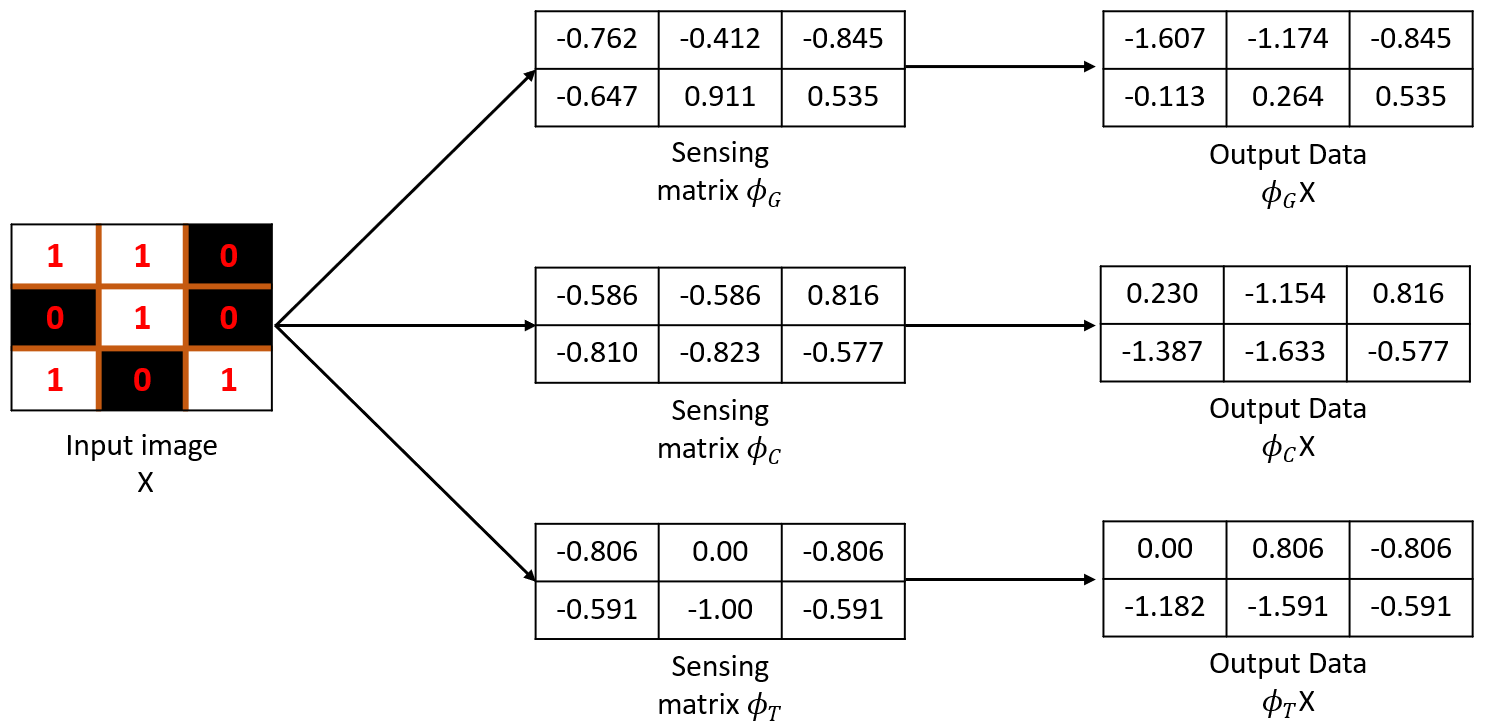}
	\caption{Feature extraction of input image using three different measurement matrices. Where $\Phi_G$  is Gaussian matrix,  $\Phi_C$  is Circulant matrix, and  $\Phi_T$ is Toeplitz matrix. The compression ratio is $30\%$.}
	\label{fig3}
\end{figure}

In this stage, the appropriate features that are required for an accurate distinguishing between infected and non-infected images will be extracted. Our method is based on the principle of CL where all features of the input image will be preserved in low dimension representation. Since the compressing procedure is done using a sensing matrix, we claim that different sensing matrices will hold the original features in different weights. Fig. \ref{fig3} shows an example for applying different measurement matrices on an image. 

It is obviously that the image has been represented in a new domain differently in every single matrix. The authors in \cite{awedat2020} have addressed this issue and proved that the classification performance of the classifier is varied based on the sensing matrix and the compression ratio. In our technique, we went further and stated that the selected features for classification would be composed from three different manipulation matrices. The input features to the classifier contain three channels and every channel is a representation of images under one sensing matrix $\Phi$. Each image has been represented by one channel. The three channels produce concatenating image under $ \Phi_1$,$\Phi_2$, and $ \Phi_3$. These matrices could be any combination of sensing matrices that proposed or applied for compressing sensing. Basically, the raw data has been directly compressed and forwarded as features for classification purpose. In this study, we applied Gaussian matrix, Circulant matrix, and Toeplitz matrix. This selection is not unique, but it is just to confirm the effectiveness of our technique. 

\subsection{Data Classification}
Once the features of input CT images have been selected, the next step in our method is the classification process. We have implemented deep learning network with several convolution and pooling layers. Then feed-forward back-propagation method is used for the learning rule selection to extract the features to classify the positive and negative COVID-19 samples. The Max pooling is utilized with kernel size of $2\times 2$ to minimize the size of the convolved features. There is no need to resize the input feeding images to the classifier since they are already compressed to the required size using sensing matrix.

\section{Experimental setup}
The proposed CADe system is developed using deep learning based on compressing learning models for classification of the raw data without any kind of preprocessing. The implementation process was conducted using Python programming language on 24 Intel(R) Xeon(R) CPU E5-4607 0 @ 2.20GHz, 377G memory and two Quadro P2000.

\subsection{Dataset (CT Images)}
\begin{figure}[h]
	\centering
	\hspace*{-.2cm}
	\includegraphics[width=.48\textwidth,height=5.1cm]{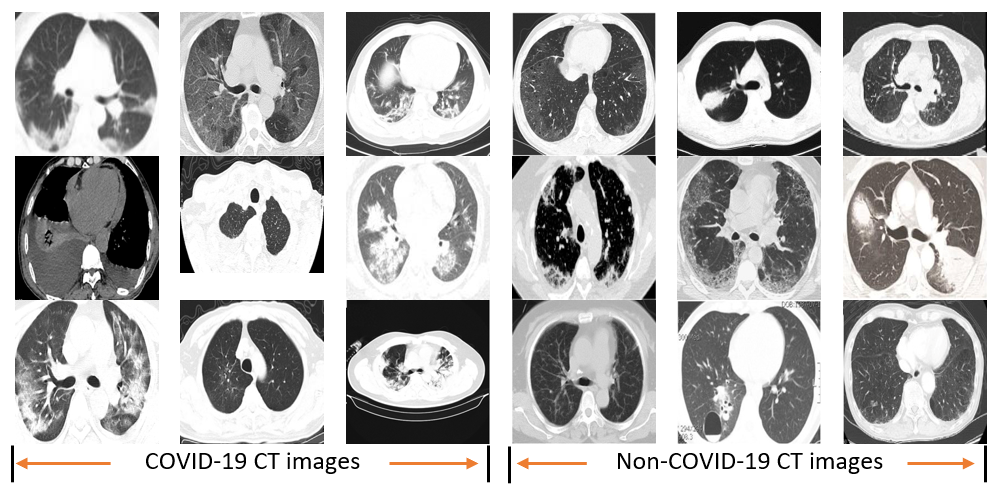}
	\caption{CT positive and negative samples from the dataset for COVID-19 diagnosis.}
	\label{fig4}
\end{figure}

The COVID-CT dataset that has been used in this study is is publicly available \cite{yang2020covid}. There are 349 images of COVID-19 collected from 216 patients. The non-COVID-19 data contains 397 samples. The images collected from four sources MedPix, LUNA, PubMed Central (PMC), and Radiopaedia website. Fig. \ref{fig4} shows some positive and negative samples of the CT images.

In this study, the collected CT images have different sizes. Therefore, the first step was to resize the entire images into one scale. All input images have been resized to $120\times120$, which should be very efficient and accurate size for compressing images using a single sensing matrix. Since the proposed approach depends on the use of more than one sensing matrix, the entire images are represented in the grayscale domain. Then the three sensing different matrices are set to have the same size, which accordingly produces output compressed images with final size of $64\times64$.

\subsection{Results }
\begin{figure}[h]
	\centering
	\hspace*{-.2cm}
	\includegraphics[width=.45\textwidth,height=6.9cm]{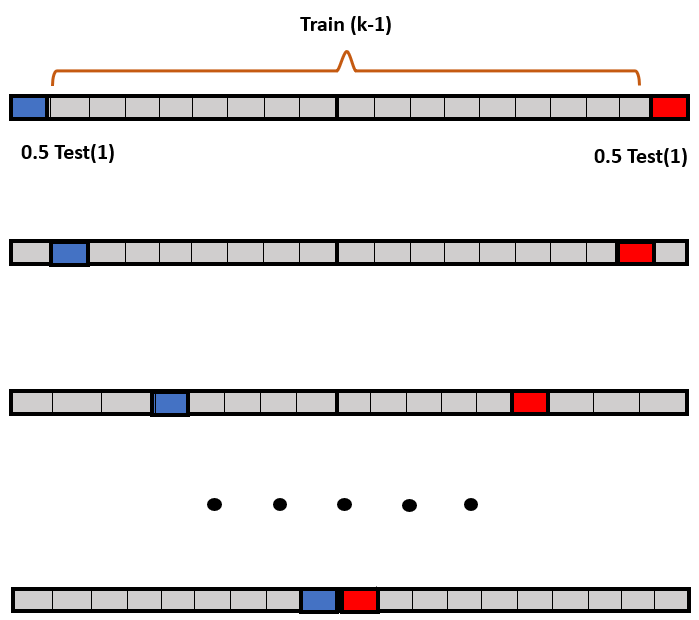}
	\vspace*{2.2mm}
	\caption{K-fold cross validation for the input CT images. Every fold contains the same number of images from each class.}
	\label{fig5}
\end{figure}

After the model has been built successfully and to avoid any bias, the dataset used was randomly split into two independent parts training and testing respectively. Then K-fold cross validation method was applied to obtain several results according to each observation from the raw dataset, which means each sample could be considered in both cases training set and testing set. After successfully training the model, we have divided the testing images equally between two categories at each fold. The positive and negative COVID images are randomly mixed. Fig. \ref{fig5} displays how testing and training sets are selected. 

All trained models are evaluated using the accuracy and validation loss (val-loss). The starting point is that testing the CL technique for the classification. We have applied three sensing matrices Gaussian, Circulant, and Toeplitz to manipulate the images into the size of $64\times64$. Then apply a quick comparison with the original images where there is not any kind of compression sensing (No CS). In the classifier, the original images resized to the same size of compressed images. The CT images dataset was randomly split into two independent parts with $80\%$ and $20\%$ for training and testing respectively. The quantitative results based on k-fold cross validation method and according to 5 different k values $(k = 1-5)$ show around $86.08\%$ testing accuracy on the overall completely different testing samples. Table \ref{table1} shows the experimental result comparisons where it is obviously when the CL has been applied the classification accuracy is improved at all three different matrices comparing to the case of no compression sensing is applied.
\begin{table}[ht]
\centering
\caption{Comparison of non-compression sensing and compression sensing with different compressing methods.} \label{table1}
\includegraphics[width=8cm,height=2.8cm]{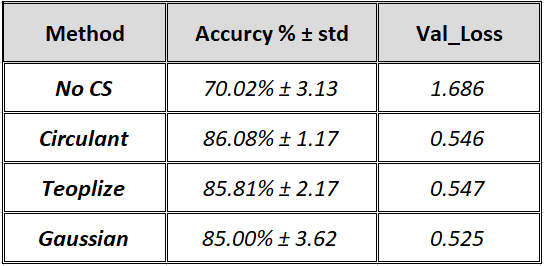}
\end{table}

The experimental results also show that the classifier performance can be improved with a minimum margin around $15\%$, which means the evaluation parameters would be affected by the compression sensing method. As we can see that Circulant matrix outperforms other matrices with a small margin around $1.5\%$. Notice that the input features to the classifier are different from three sources of compression.

\begin{figure}[h]
	\centering
	\hspace*{-.2cm}
	\includegraphics[width=.47\textwidth,height=6.7cm]{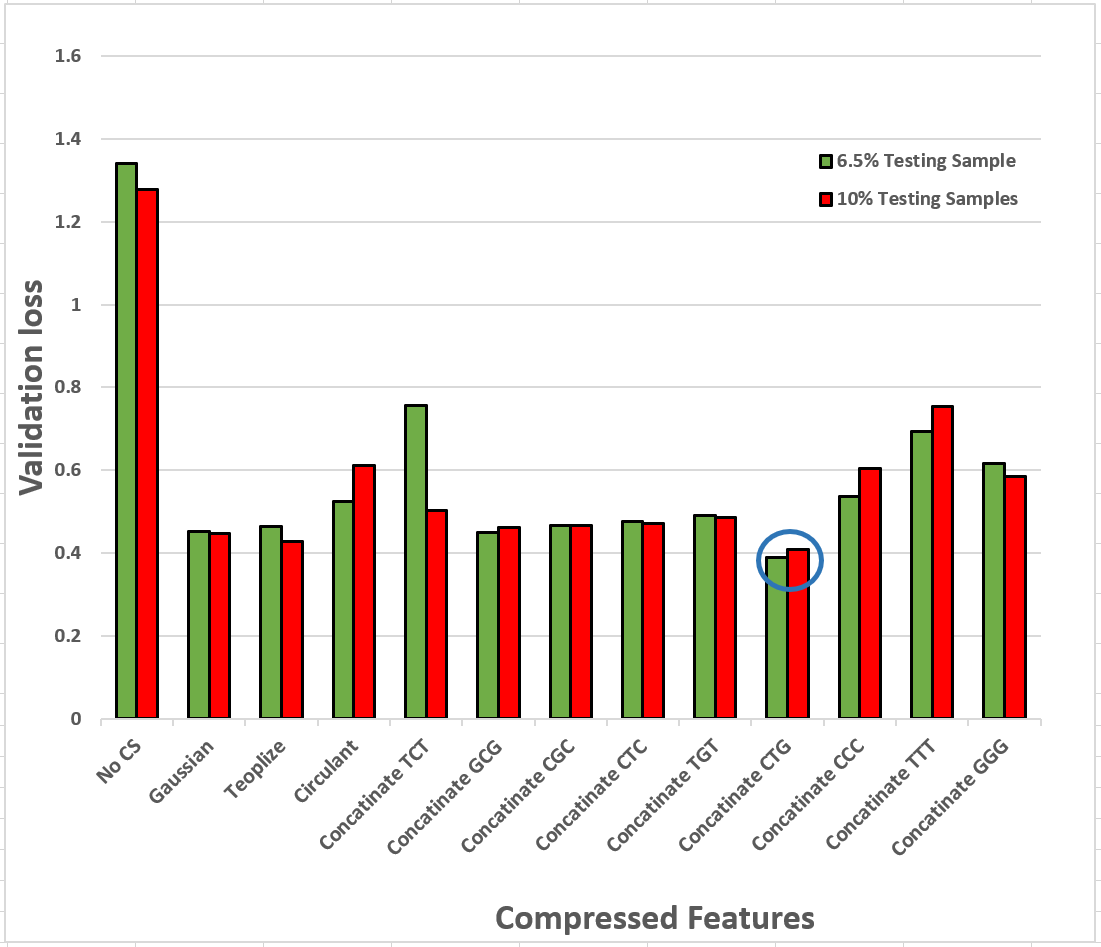}
	\vspace*{2.2mm}
	\caption{The validation loss comparison for different combination methods.}
	\label{fig6}
\end{figure}

In addition, for quantitative justification after confirming that the CL can be involved to improve the performance of the classifier, we have investigated the combination of these three different methods for classification extensively. As shown in Fig. \ref{fig2}, the three compressed features from every image are concatenated into one channel. Every channel has a size $64\times64$. First, we investigate all concatenation options to identify which one provides the best performance. Then in the second part, we compare our results with other methods that have been listed in \cite{yang2020covid}. Table \ref{table2} and Fig. \ref{fig6} show the experimental results for two testing sets. In the first set, we have left $6.5\%$ of the data samples for testing and have used the rest training. While in the second set, we have utilized $10\%$ for testing and the rest for training.
 
In general, the quantitative analysis that has been applied using all different concatenation options shows promising results that higher accuracy than the case of single channel method even though when the three channels are from the same sensing matrix (TTT, GGG, and CCC) Toeplitz, Gaussian, and Circulant respectively. Table \ref{table2} shows all possible combinations of these three channels. However, the best performance could be achieved when the three channels are totally different, which can be applied to define the severity of the COVID-19 disease. The main reason behind that is the selected features after representing the images in low dimensions are promoted by the combination of the three different channels. We also observed that increases the accuracy and validation loss as long as the selected features have been increased. 

\subsection{Discussion}
To evaluate our proposed method, we have made a comparison with the DenseNet-169 \cite{huang2017densely}, which has been trained under different pretraining methods according to \cite{yang2020covid}, named random initialization, Transfer learning (TL), and TL with contrastive self-supervised learning (CSSL). More details can be found in \cite{yang2020covid}. In addition, to avoid any bias, we select the same number of CT images for testing and 16-fold cross validation has been applied. 
 
The proposed COVID-CLNet detection method shows around $91.98\%$ testing accuracy whereas the highest accuracy of the comparison papers TL-CSSL shows $89.1\%$ testing accuracy. Thus, our COVID-CLNet based detection model shows around $2.88\%$ higher testing accuracy than the mentioned comparison methods. Although all these methods need some preprocessing steps such as lesion segmentation and lung mask, our proposed approach does not need any kind of preprocessing. The main observation of these qualitative results is shown in Table \ref{table3}.
\begin{table}[ht]
\centering
\caption{Performance comparison of different sensing matrices at different number of testing samples.} \label{table2}
\includegraphics[width=8.8cm,height=10cm]{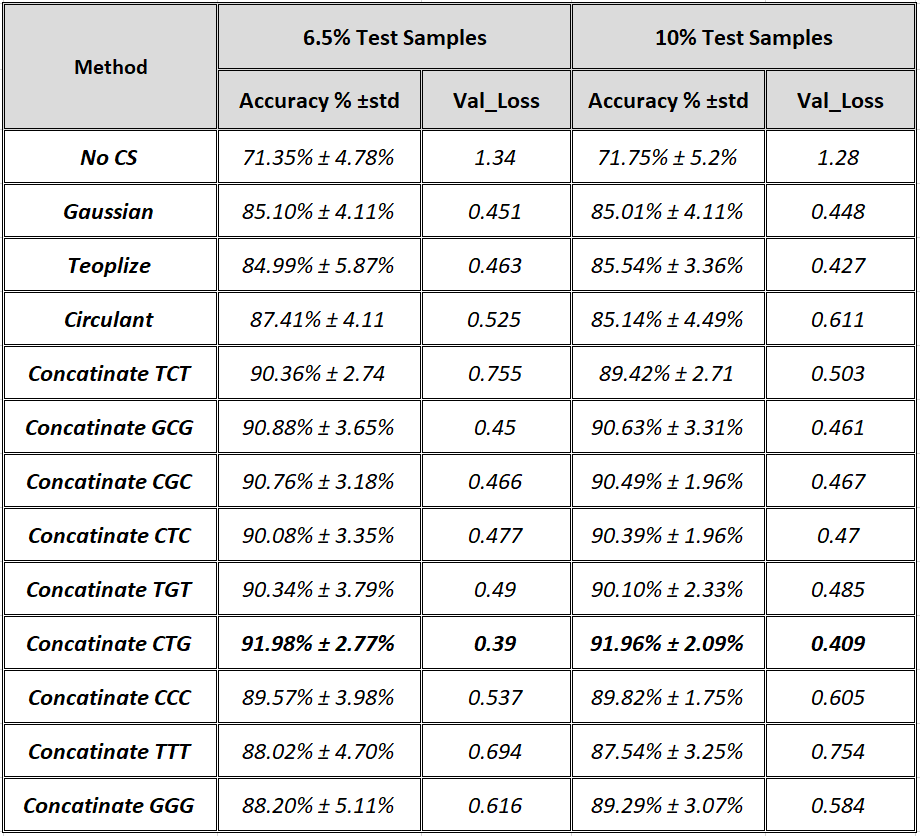}
\end{table}

\vspace*{2mm}
\begin{table}[ht]
\centering
\caption{Comparison of the proposed method with different pretrained methods.} \label{table3}
\includegraphics[width=8cm,height=1.1cm]{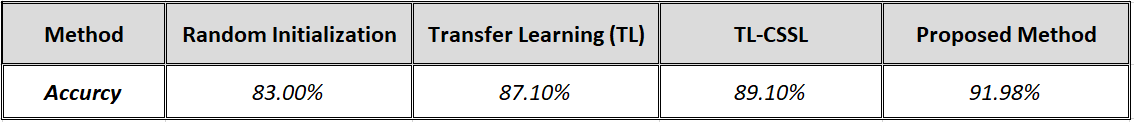}
\end{table}

\subsection{Limitations}
Even though our proposed CLNet has shown signs of high performance using raw CT images without any preprocessing, the main prevalent challenge of this work is to access a big data. Currently, most of COVID-19 datasets are limited due to the nature of the disease, patient privacy, and the requirements of the radiologist or other medical professional to data labeling. From our point of observation, data augmentation could be an option to improve the system performance and to avoid the overfitting. In our technique, there are more options to expand the dataset where augmentation could be performed on either the original dataset images or on the CL features level. 

\section{Conclusion}
The proposed CADe system for COVID-19 detection could be a great and inexpensive tool to assist the radiologists or other medical professionals to detect and identify the COVID-19 cases at early infection stages and in the shortest possible time. Our improved deep learning network model based on the compressive learning (COVID-CLNet) is applied on computed tomography (CT) images directly and without any kind of preprocessing. The observed results show very promising detection precision with $91.98 \%$ testing accuracy. The work is in progress to improve the performance of our CADe system, which could be done by either collecting more COVID-19 affected samples, applying data augmentation as preprocessing to the system, or combining the CL with different deep learning networks.

\bibliographystyle{IEEEtran}
\bibliography{REFERENCES}

\end{document}